\documentstyle[seceq,epsfig,wrapfig]{ptptex}

\def\PRL#1#2#3{Phys.~Rev.~Lett.~{\bf #1}~(#2)~#3}
\def\PRD#1#2#3{Phys.~Rev.~D{\bf #1}~(#2)~#3}

\def\PLB#1#2#3{Phys.~Lett.~B{\bf #1}~(#2)~#3}

\def\ZPC#1#2#3{Z.~Phys.~C{\bf #1}~(#2)~#3}
\def\PR#1#2#3{Phys.~Rep.~{\bf #1}~(#2)~#3}

\def\MPLA#1#2#3{{Mod.~Phys.~Lett.~}{\bf A#1}~(#2)~#3}
\def\IJMPA#1#2#3{{Int.~J.~Mod.~Phys.~}{\bf A#1}~(#2)~#3}

\def\NIMPRA#1#2#3{{Nucl.~Instr.~and~Meth.~in~Phys.~Res.~}
{\bf A#1}~(#2)~#3}

\def\beq{\begin{equation}}
\def\eeq{\end{equation}}
\def\beqa{\begin{eqnarray}}
\def\eeqa{\end{eqnarray}}

\def\mz{m_Z}

\def\ai{\alpha^{-1}}
\def\az{\alpha(\mz)}
\def\aiz{\alpha^{-1}(\mz)}
\def\da{\Delta\alpha}
\def\das{\Delta\alpha(s)}
\def\daz{\Delta\alpha(\mz)}

\def\dahs{\Delta\alpha^{(5)}_{hadrons}(s)}
\def\dahz{\Delta\alpha^{(5)}_{hadrons}(\mz^2)}

\def\as{\alpha_s}

\def\asz{\alpha_s(\mz)}
\def\seff{\sin^2\theta_{\rm eff}^{\rm lept}}

\def\vev#1{\langle #1 \rangle}
\def\f{f}

\notypesetlogo  %comment in if to eliminate PTPTeX logo

\preprintnumber{CERN--TH/96--79\\hep--ph/9603415}

\title{\boldmath
The Status of the Determination of $\az$ and $\asz$.
}

\author{
Tatsu {\sc Takeuchi}\footnote{
Current address: TH~Division, CERN,  
CH--1211~Gen\`eve~23, Switzerland. \\ 
E-mail address: takeuchi@vxcern.cern.ch}
}

\inst{
Fermi~National~Accelerator~Laboratory \\
P.~O.~Box 500, Batavia, IL 60510, U.S.A.
}

\abst{
I will discuss the current status of the determination 
of $\az$ and $\asz$, emphasizing the pitfalls that one must
avoid in performing statistical analyses. 
}

\begin{document}
%
% Title page for CERN preprint
%
\pagestyle{empty}
\begin{flushright}
{CERN--TH/96--79\\hep--ph/9603415}
\end{flushright}
\vspace*{3cm}
\begin{center}
{\Large\bf\boldmath The Status of the Determination of $\az$ and $\asz$} \\
\vspace*{1.5cm}
{\large\bf Tatsu Takeuchi} \\
\vspace{0.3cm}
{\it TH Division, CERN, CH--1211 Gen\`eve 23, Switzerland} \\
\vspace*{5cm}
{\bf ABSTRACT} \\ \end{center}
\vspace*{5mm}
\noindent
I will discuss the current status of the determination 
of $\az$ and $\asz$, emphasizing the pitfalls that one must
avoid in performing statistical analyses. 
\vspace*{2cm}
\begin{center}
{\it Talk presented at the Yukawa International Seminar (YKIS) '95 \\
     ``From the Standard Model to Grand Unified Theories'' \\
     Kyoto, Japan, August 21--25, 1995.}
\end{center}
\vfill
%\vspace*{2cm}
\begin{flushleft} 
CERN--TH/96--79 \\
March 1996
\end{flushleft}
\eject
%
% beginning of PTPTEX file
%
\setcounter{page}{1}
\pagestyle{myheadings}

\maketitle

\section{Introduction}

Precise determinations of the values of 
$\az$ and $\asz$ are important in the context of both
the Standard Model (SM) and Grand Unified Theories (GUT's)
which are the two major themes of this conference.

In the SM, the values of $\az$ and $\asz$ are necessary as
inputs in calculating the predictions for observables at the
$Z$--mass scale.   The accuracy in which they are known is
reflected directly onto the accuracy of the SM predictions
that must be compared with the precision electroweak
measurements at LEP/SLC. 
Therefore, if one wishes to test the SM at the level of
radiative corrections and/or detect virtual effects of
new physics in the LEP/SLC measurements,
one needs to know $\az$ and $\asz$ to very high accuracy.  
This point is well summarized in the talk by 
Hagiwara in this proceedings \cite{HAGIWARA}.

In the case of GUT's, the three gauge couplings
of $SU_C(3)\times SU_L(2)\times U_Y(1)$ must
coincide when evolved up to the GUT scale using 
renormalization group equations (RGE's).  
Whether they actually coincide or not depends on their
initial values at $s=\mz^2$, and the particle spectrum 
contributing to the RGE's.
Therefore, precise values of $\az$ and $\asz$ will
provide important constraints on the particle content
and their masses for any given GUT \cite{GUTS}.

In this talk, I will discuss the current status of
the determination of $\az$ and $\asz$.
To avoid this talk from being a simple update of
Ref.~\citen{UCLA95}, I have shifted my emphasis somewhat
to the problems in statistical analyses that one encounters
when trying to extract these numbers from experimental data.

For $\az$, I will give a comprehensive review of
all the recent attempts to determine its value from the 
$\sigma(e^+e^-\rightarrow hadrons)$ data
\cite{BJPV,JEGER,MORRIS1,MARZEP,EJ,BP,MORRIS2} and clarify where 
these analyses differ from each other.  
In particular, I will emphasize 
the problems that must be addressed in the treatment of
the experimental data,
and whether any of these analyses
have succeeded in doing so or not.
My conclusion here will be somewhat different from the
one I gave at the conference, partly due to the fact that the
details of the Swartz 1995 analysis has been made 
available\cite{MORRIS2} and it can now be given a proper 
evaluation, and partly due to my better understanding of the
problem.

For $\asz$, I will not be reviewing all 
the different methods of determining $\asz$ and how 
reliable they are, since 
thorough and excellent reviews on the subject 
by experts in the field already exist in the literature
\cite{HINCH,SHIFMAN}.  
I will instead focus on the problem of determining $\asz$
from the $Z$--lineshape parameter 
$R_\ell = \Gamma(Z\rightarrow hadrons)/\Gamma(Z\rightarrow \ell^+\ell^-)$
at LEP and discuss
an interesting (but erroneous) speculation of Ref.~\citen{CONFER} 
which claims that the error on $R_\ell$, 
and thus $\asz$ determined from it, may be 
grossly underestimated based on an exotic 
(to particle physicists) statistical analysis.

\section{\boldmath The value of $\az$}

\begin{figure}[t]
\begin{center}
\epsfig{file=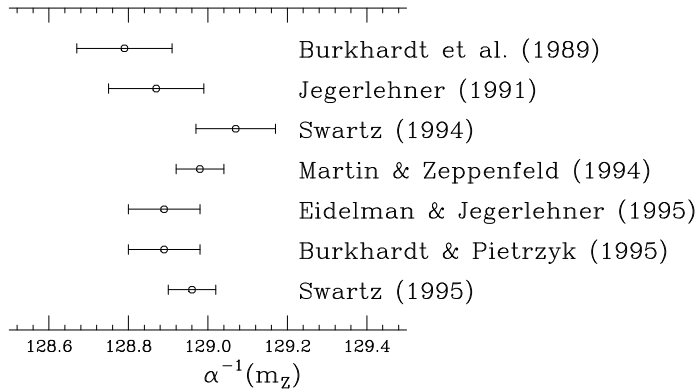,angle=90,height=2in}
\caption{
Evaluations of $\az$ by different authors.
All values have been rescaled to $\mz=91.1887$ GeV and
the top quark contribution has been removed.
}
\label{azlist}
\end{center}
\end{figure}

The best determination of
$\az$ until late 1994 had been that of Jegerlehner\cite{JEGER} 
from 1991 which was
\beq
\aiz = 128.87 \pm 0.12.
\eeq
As an example of how this uncertainly in $\az$ affects SM
predictions, I quote the calculation of
$\seff$ based on this value using the FORTRAN program 
ZFITTER~4.9\cite{ZFITTER} which gives
\beq
\seff = 0.2319 \pm 0.0003
\eeq
for the SM with $m_t = 180$GeV and $m_H = 300$GeV.
The 0.1\% error on $\az$ has propagated directly into
a similar error on $\seff$.  On the other hand, the
experimental error on $\seff$ has been decreasing steadily. 
Currently, the averaged value of $\seff$ 
from all asymmetry measurements at LEP and SLC is at\cite{LEPDATA}
\beq
\seff = 0.23143 \pm 0.00028.
\eeq
Clearly, an improvement on the knowledge of $\az$ is called
for given that the experimental error is now smaller than the
theoretical error, and can be expected to decrease even further.

During the past year or so, several authors have independently made
attempts to reevaluate the value of $\az$ through careful
reanalyses of existing $\sigma(e^+e^-\rightarrow hadrons)$ data
\cite{MORRIS1,MARZEP,EJ,BP,MORRIS2}.
The hope was that the uncertainty in $\az$ could be reduced
since some new data on $\sigma(e^+e^-\rightarrow hadrons)$ had
been released\cite{ND,MD1,DM2}, and the 
$O(\alpha_s^3)$ QCD correction to the cross section\cite{KATAEV} 
together with a better determination of 
$\asz$\cite{HINCH,SHIFMAN} was now
available.
The resulting new values of $\az$ are shown in Fig.~\ref{azlist}
together with a couple of older values.

Comparing the five new evaluations,
we see that the four most recent ones 
are all more or less consistent with each other and the
Jegerlehner '91 value with no significant decrease in the error.
The only number which is slightly off is the 
Swartz '94 value from Ref.~\citen{MORRIS1} which was updated by 
Swartz himself in Ref.~\citen{MORRIS2}.

Due to this agreement between different evaluations, 
it does not really matter which of these four numbers is used 
as the standard value of $\az$ from a practical point of view.
However, since we must decide on one number anyway, 
we may as well choose the one which can be considered 
to give the best estimate of the true value of $\az$.

In the following, I will review how the analyses
shown in Fig.~\ref{azlist} differ from each other so that we may
make an informed decision.
But before I begin, I will first review the definition
of $\az$ and how it can be determined from the 
$\sigma(e^+e^-\rightarrow hadrons)$ data so that we
can have a clear understanding of the issues that must be
addressed.

\subsection{The Definition of $\az$}

The quantity whose value is usually quoted as that of the
``effective QED coupling constant at the $Z$ mass scale'' 
is defined as
\beq
\aiz = \ai\left[ 1 - \daz 
          \right]
\eeq
with
\beq
\das = 4\pi\alpha 
        {\rm Re}\left[ \Pi'_{QQ}(s) - \Pi'_{QQ}(0)
                \right].
\eeq
Here, $\Pi'_{QQ}(s)$ is the photon vacuum polarization function
defined as
%
%\beqa
%\int d^4x\,e^{iq\cdot x}\langle J_Q^\mu(x) J_Q^\nu(0)
%                        \rangle
%& = & i\left( g^{\mu\nu} - \frac{q^\mu q^\nu}{q^2}
%       \right) \Pi_{QQ}(s) \cr
%& = & i(q^2 g^{\mu\nu} - q^\mu q^\nu)\Pi'_{QQ}(s)
%\eeqa
\beq
\int d^4x\,e^{iq\cdot x}\langle J_Q^\mu(x) J_Q^\nu(0)
                        \rangle
= i(q^2 g^{\mu\nu} - q^\mu q^\nu)\Pi'_{QQ}(s)
\eeq
where $J_Q^\mu(x)$ is the electromagnetic current modulo the
coupling constant $e$.

It has been customary to include 
only the {\it light fermion} contributions to $\daz$.
The top contribution had been excluded from $\daz$
because the top mass was unknown until quite recently\cite{TOP}
(though care is needed when comparing results 
since recent authors include it) and the
$W$ contribution was also excluded to keep the definition of
$\az$ gauge independent.  They are also numerically small 
compared to the light fermion contribution since the
$W^+W^-$ and $t\bar{t}$ thresholds are above the $Z$ mass so that
they do not contribute logarithms to the running of 
$\alpha(s)$ between $s=0$ and $s=\mz^2$.
I will adhere to this convention in this talk and only discuss the
contribution of light fermions.
Readers who are interested in how the value of
$\az$ will change when the $t\bar{t}$ and the
$W^+W^-$ {\sl pinch} contributions are included are referred to
Table~I of the talk by Hagiwara\cite{HAGIWARA}.

\subsection{Contribution of the light fermions}

The contribution of the leptons to $\da$ can be calculated
accurately in perturbation theory and one finds
\beqa
\da_{leptons}(\mz^2)
& = & \sum_{\ell=e,\mu,\tau}
      \frac{\alpha}{3\pi}
      \left[ - \frac{8}{3} + \beta_\ell^2
             - \frac{1}{2}\beta_\ell(3 - \beta_\ell^2)
               \ln\left( \frac{1-\beta_\ell}{1+\beta_\ell}
                  \right)
      \right]   \cr
& = &
      \sum_{\ell=e,\mu,\tau}
      \frac{\alpha}{3\pi}
      \left[ \ln\left( \frac{\mz^2}{m_\ell^2}
                \right)
           - \frac{5}{3}
           + O\left( \frac{m_\ell^2}{\mz^2}
              \right)
      \right]    \cr
& = & 0.03142,
\eeqa
where $\beta_\ell = \sqrt{1 - 4m_\ell^2/\mz^2}$.

On the other hand,
the contribution of the five light quarks ($u,d,s,c,b)$ to $\da$
cannot be calculated perturbatively.  Instead,  
unitarity and the analyticity of $\Pi'_{QQ}(s)$ is used to write
\beq
\dahs
 = \frac{\alpha s}{3\pi}{\rm P}
   \int_{4m_\pi^2}^\infty ds' \frac{R(s')}{s'(s-s')},
\label{INTEG}
\eeq
where
\beq
R(s) \equiv  
\frac{3s}{4\pi\alpha^2(s)}
\sigma(e^+e^- \rightarrow \gamma^* \rightarrow hadrons)
= -12\pi{\rm Im}\Pi'_{QQ}(s)
\eeq
and the functional form of $R(s)$ is extracted from experiment.
It is this reliance on the experimental values of $R(s)$,
which are currently accompanied by large experimental errors, 
that we end up with a relatively large error on $\az$ even though the
fine structure constant $\alpha$ is known to extreme accuracy.

\subsection{Problems with using $R(s)$}

There are a couple of problems that must be addressed when using 
the experimental values of $R(s)$ to calculate $\az$.

First, the data for $R(s)$ are only available for discrete,
scattered values of $s$. One must therefore 
interpolate between, and extrapolate beyond the available data points
to make use of Eq.~\ref{INTEG}.  

Two methods have been used in the literature to deal with this problem.
The first is to connect the data points directly with straight lines
and perform trapezoidal integration\cite{BJPV,JEGER,EJ},
and the second is to guess the functional form of $R(s)$ and
fit it to the data\cite{BJPV,MORRIS1,MARZEP,MORRIS2}.

Both methods have their pros and cons.  Trapezoidal integration
is free from human prejudice about the functional form of
$R(s)$ but it is difficult to take into account the experimental
errors properly: sparsely distributed precise data points may not
get the appropriate weight relative to the densely spaced 
data points with larger errors, and possible   
correlations between errors are not accounted for.
Connecting two data points that are far apart with a straight 
line will also introduce errors in regions where $R(s)$ is 
changing rapidly. (A kind of `human prejudice' in a sense.) 

On the other hand, fitting a guessed functional form to the data
has the advantage that experimental errors are easier to take into
account (though care is needed in treating 
normalization errors\cite{DAGOSTINI} as will be discussed later).
However, the result will depend on the choice of the fit function
and its parameterization, and
systematic errors and biases will be introduced
that are difficult to estimate.

The two methods are often combined 
({\it e.g.}\ fitting the Gounaris--Sakurai\cite{GS} 
form to the $\rho$\cite{KINOSHITA},
Breit--Wigner forms to the narrow resonances,
and using trapezoidal integration for the continuum) and  
are supplemented by the use of perturbative QCD for
the high energy tail of $R(s)$.

Secondly, the experimental values of $R(s)$ are 
always accompanied by experimental errors, 
both statistical and systematic, and often
rather large.   

The systematic error consists of uncertainties in the
measurement/monitoring of collider luminosity ${\cal L}$,
Monte Carlo calculation of the detection efficiency $\varepsilon$,
and radiative corrections $\delta$
that are introduced when converting the
measured event rate $N$ into the cross section $\sigma$ 
through the relation
\beq
N = \sigma{\cal L}\varepsilon(1+\delta).
\eeq
It is therefore a {\it normalization uncertainty}.

Obviously, such normalization errors on data points
belonging to the same experiment are 100\% correlated.
This type of correlation of systematic errors within a single 
experiment is called a Type~I correlation in Ref.~\citen{MORRIS2}.

However, there can also be strong correlations between the 
normalization errors on data points belonging to two different 
experiments if they use the same collider, 
same knowledge of radiative corrections, 
event generators based on similar models,
same background estimation methods, same detector response models, etc.
This type of correlation of systematic errors between
different experiments is called a Type~II correlation in 
Ref.~\citen{MORRIS2}.
Their exact size is difficult to estimate, 
but they nevertheless exist.

Taking such correlations into account
when extracting $R(s)$ from data has important 
consequences as we will see later.

\subsection{Recent estimates of $\az$}

Let us now start our tour of all the different
analyses shown in Fig.~\ref{azlist} and see how they 
address the problems discussed above.
The order in which they will be presented is not
necessarily chronological, but what
I deem appropriate to contrast the differences.

\subsubsection{The analysis of Burkhardt et al.}

At the time when LEP started running in 1989, the 
most accurate determination of $\dahz$ was that given by
Burkhardt et al. in Ref.~\citen{BJPV}:
\beq
\dahz = 0.0286 \pm 0.0009. 
\label{azBURK}
\eeq
(Actually, Ref.~\citen{BJPV}  reports the
value of $\dahs$ at $\sqrt{s} = 92$ GeV.
Rescaling to $\sqrt{s} = \mz = 91.1887$ GeV gives the
above value\cite{LEP1}.)
This value was calculated using the 
following three methods of interpolation
\begin{enumerate}
\item{Trapezoidal integration for the continuum and the $\rho$. \\
      Breit--Wigner forms for the narrow resonances.}
\item{Trapezoidal integration for the continuum 
      after a partial smoothing out of the $R(s)$ data. 
      (Ref.~\citen{BJPV} does not give any detail on 
       how the smoothing was done.)\\
      Breit--Wigner forms for the narrow resonances.}
\item{Breit--Wigner forms for the $\rho$ and narrow resonances.  \\
      Linear interpolation in $W=\sqrt{s}$ for every few points in the
      continuum.}
\end{enumerate}
In all three cases, 
perturbative QCD (at $O(\as^2)$) was used for $R(s)$ above $40$ GeV.
The difference in $\dahz$ due to the difference in the method of
interpolation was found to be negligible
compared to the statistical error.

Eq.~\ref{azBURK} corresponds to
\beqa
\aiz & = & \ai \left[ 1 - \da_{leptons}(\mz^2) - \dahz
               \right]     \cr
     & = & 128.81 \pm 0.12
\label{az89}
\eeqa
and results in an error of about 0.1\% in $\az$.
The region below the bottom threshold contributed about
$1/3$ of $\dahs$, and 80\% of the error.

\begin{figure}[t]
\begin{center}
\epsfig{file=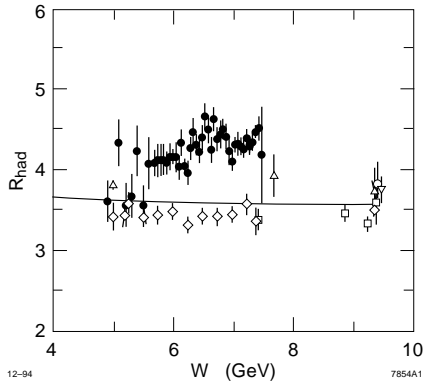,angle=0,height=2in}
\caption{Comparison of MARK~I (black circles) and 
Crystal Ball (while diamonds) data.
Only statistical errors are shown.
The solid line shows the perturbative QCD result.}
\label{MKIvsCB}
\end{center}
\end{figure}

\subsubsection{The analysis of Jegerlehner}

The result of Ref.~\citen{BJPV} was subsequently updated in 1991
by Jegerlehner\cite{JEGER}, who was one of the original authors,
in which the data from the MARK~I collaboration\cite{MARKI} 
in the energy region 5--7.4 GeV were replaced by the
more accurate data from the Crystal Ball collaboration\cite{CBALL}.
Fig.~\ref{MKIvsCB} shows the data of both collaborations in the energy range in
question, together with the perturbative QCD result.
The method of integration was the same as method No.~1 of Ref.~\citen{BJPV}
and the result was
\beq
\dahz  =  0.0282 \pm 0.0009,
\eeq
with no change in the size of the error.
This corresponds to
\beq
\aiz  =  128.87 \pm 0.12     
\label{JEGERVALUE}
\eeq
which had been the standard value until late 1994.

\subsubsection{The analysis of Eidelman and Jegerlehner}

The 1991 Jegerlehner value was again updated by
Eidelman and Jegerlehner in Ref.~\citen{EJ} to include
new data\cite{ND,MD1,DM2} in the analysis.
Again trapezoidal integration was used.

This reference goes into some detail on how 
experimental errors were taken into account.
In order to prevent sparsely distributed precise data points
from getting less weight than densely spaced data points with
large errors, which is one of the problems inherent in
trapezoidal integration, Eidelman and Jegerlehner used the following two
methods:
\begin{itemize}
\item{
When more than one experiment gave results in the
same energy region, the contribution to $\dahz$ from that energy region
was calculated using trapezoidal integration for each experiment
separately, and then the resulting integrals were combined by taking a
error weighted average.
}
\item{
Data from different experiments in the same energy region were
combined by taking local error weighted averages, 
and trapezoidal integration was performed on the combined data.
}
\end{itemize}
Both methods were found to give similar results
which is not too surprising since the
local average for a point $s$ was defined as
the average of the $R(s)$ values obtained by linear
interpolation between the nearest data points.

After subtracting out the top quark contribution from the value
reported in Ref.~\citen{EJ}, we obtain
\beq
\dahz = 0.02804 \pm 0.00065,
\eeq
and
\beq
\aiz = 128.89 \pm 0.09,
\eeq
which agrees very well with the previous estimate\cite{JEGER}.

\subsubsection{The analysis of Burkhardt and Pietrzyk}

The recent analysis of Burkhardt and Pietrzyk in Ref.~\citen{BP} is
another update of Ref.~\citen{BJPV}, but this time
using method No.~3, i.e. linear interpolation in $W=\sqrt{s}$ 
for every few points in the continuum.

In the $5-7.4$ GeV energy range, an error weighted average of
the MARK~I\cite{MARKI} and Crystal Ball\cite{CBALL} data
was used instead of discarding the MARK~I data altogether as
in Refs.~\citen{JEGER,MORRIS1,EJ,MORRIS2}.
New data from Refs.~\citen{ND,MD1,DM2} were also included.
The resulting value of $\dahz$ was
\beq
\dahz = 0.0280 \pm 0.0007,
\eeq
which corresponds to
\beq
\aiz = 128.89 \pm 0.09.
\eeq

\subsubsection{The analysis of Martin and Zeppenfeld}

The analysis of Martin and Zeppenfeld in Ref.~\citen{MARZEP}
distinguishes itself from all the other analyses in its extensive use of
perturbative QCD.   In table~\ref{comp1}, I list the energy ranges in which
different authors have applied perturbative QCD.
(Though I also list the values of $\asz$ that were used,
at the current level of accuracy, the difference is insignificant.
Changing the value of $\asz$ has little effect on the resulting value
of $\az$.)

\begin{figure}[t]
\begin{center}
\epsfig{file=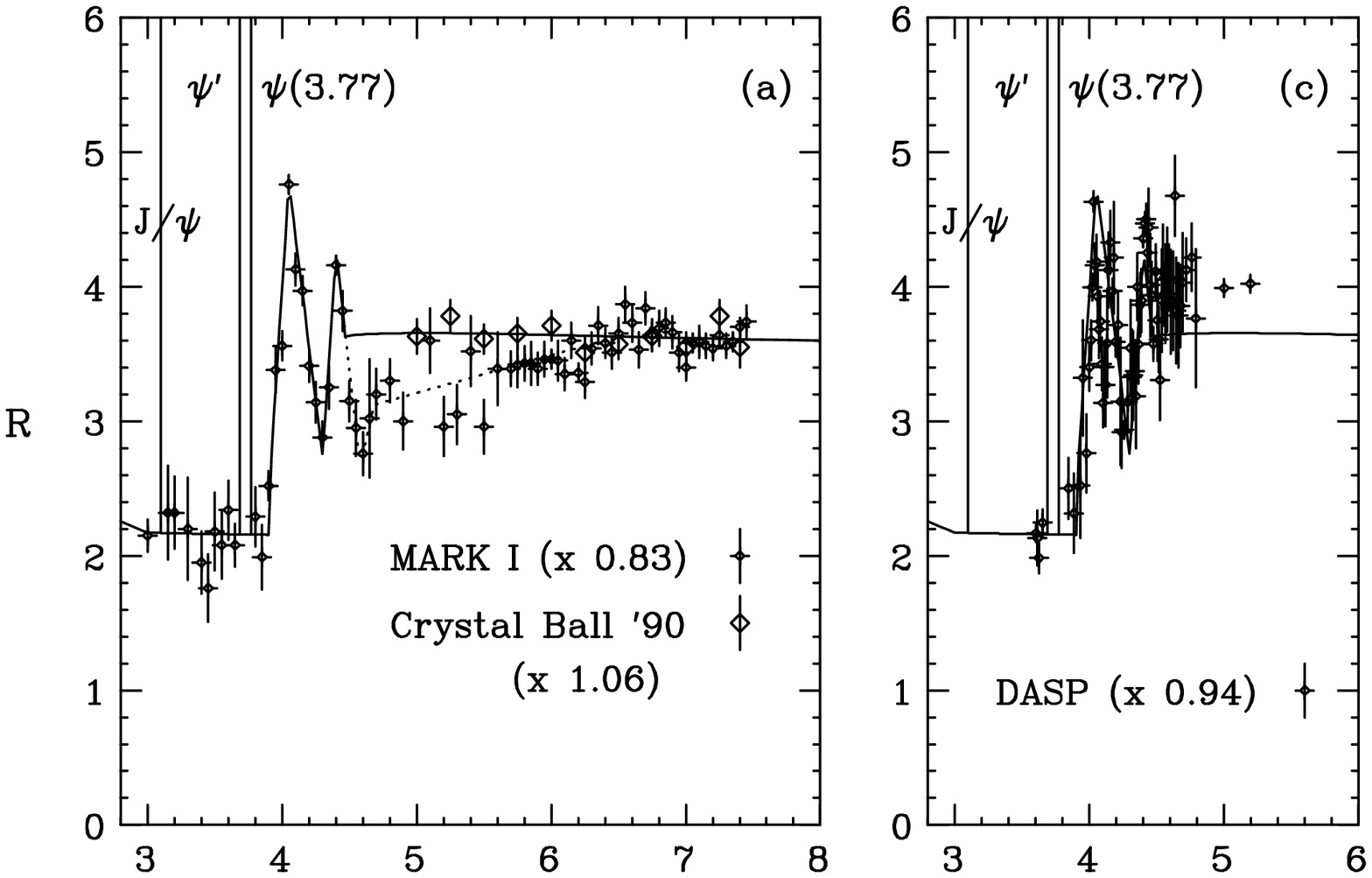,angle=90,width=4.5in}
\epsfig{file=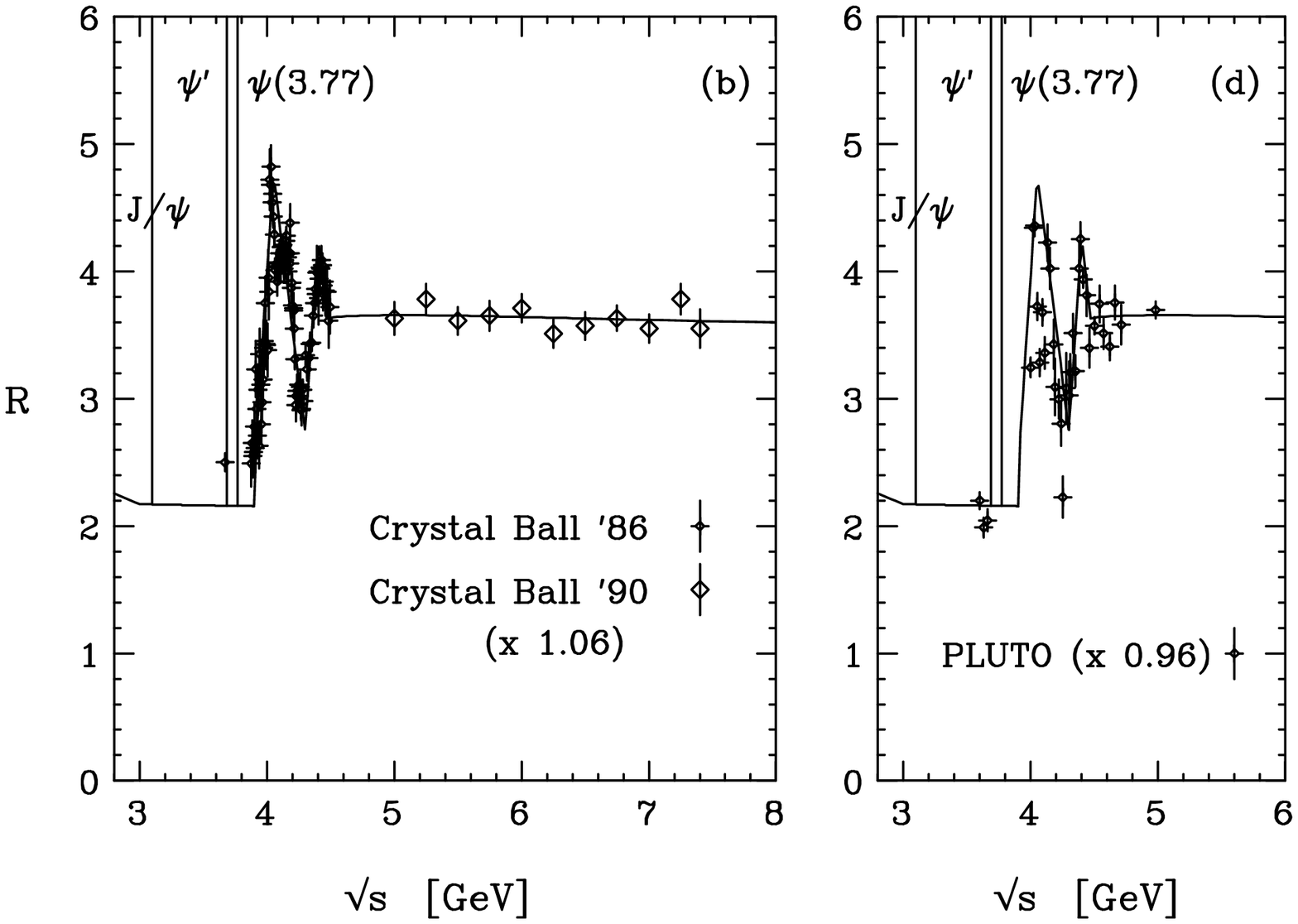,angle=90,width=4.5in}
\caption{The rescaling of data in Ref.~\protect\citen{MARZEP}.
Only statistical errors are shown.}
\label{MZFIG}
\end{center}
\end{figure}

In the two energy regions 3--3.9 GeV and 6.5--$\infty$ GeV,
Martin and Zeppenfeld express $R(s)$ as the perturbative QCD value
plus the $J/\Psi$, $\Psi$, and $\Upsilon$ resonances.
The DASP\cite{DASP}, PLUTO\cite{PLUTO}, MARK~I\cite{MARKI}, and
Crystal Ball\cite{CBALL} data are all rescaled to fit the perturbative
QCD result in these energy ranges, and the rescaled data is used
to resolve a couple of resonances in the
energy region between 3.9 and 6.5 GeV.  See Fig.~\ref{MZFIG}.

This kind of collective rescaling of data
is not as arbitrary as it may seem.  
The normalization errors reported by the experiments 
shown in Fig.~\ref{MZFIG} are
15\% (DASP), 12\% (PLUTO), 10$\sim$20\% (MARK~I), 
and 5.2\% (Crystal Ball), respectively, so the rescaled
data are well within $1\!\sim\! 2\sigma$ of their original
values. 
If one believes that perturbative QCD is valid in this
energy range, then it is quite natural to use it to fix 
the data to the `correct' normalization. 

\begin{table}[t]
\begin{center}
\caption{Comparison of the reliance on perturbative QCD
in the evaluation of $\az$ between different authors.
}
\label{comp1}
\begin{tabular}{|l|c|c|c|}
\hline
Author & Energy Range (GeV) & Order in $\as$ & $\asz$ \\
\hline
Burkhardt et al.\cite{BJPV,LEP1} (1989) 
& $40-\infty$
& $2$
& $0.12 \pm 0.02$ \\
Jegerlehner\cite{JEGER} (1991)
& $40-\infty$
& $2$
& $0.117 \pm 0.010$ \\
Swartz\cite{MORRIS1} (1994)
& $15-\infty$
& $3$
& $0.125 \pm 0.005$ \\
Martin \& Zeppenfeld\cite{MARZEP} (1994)
& $3-3.9$, $6.5-\infty$
& $3$
& $0.118 \pm 0.007$ \\
Eidelman \& Jegerlehner\cite{EJ} (1995)
& $40-\infty$
& $3$
& $0.126 \pm 0.005$ \\
Burkhardt and Pietrzyk\cite{BP} (1995)
& $12-\infty$
& $3$
& $0.124 \pm 0.020$ \\
Swartz\cite{MORRIS2} (1995)
& $15-\infty$
& $3$
& $0.116 \pm 0.005$ \\
\hline 
\end{tabular}
\end{center}
\end{table}

Due to the heavy reliance on perturbative QCD in this analysis, 
and consequently the relatively light reliance on experimental data,
the uncertainty in $\dahz$ is reduced.
The value quoted in Ref.~\citen{MARZEP} is
\beq
\dahz = 0.02739 \pm 0.00042,
\eeq
which corresponds to
\beq
\aiz = 128.98 \pm 0.06.
\eeq

\subsubsection{The analyses of Swartz}

The analyses by Swartz in Refs.~\citen{MORRIS1} and \citen{MORRIS2} 
fit a smooth function described by polynomials in 
$W = \sqrt{s}$ to the continuum part of $R(s)$.
Resonances were described by Breit--Wigner forms as usual, 
and perturbative QCD (at $O(\as^3)$) was used above 15 GeV.

The Swartz analyses distinguish themselves as the first
attempts to treat Type~I and Type~II correlations on a firm
statistical basis. 

In his first analysis\cite{MORRIS1}, 
Swartz used a standard technique in fitting his function 
$R_{fit}(s;a_k)$ to the $R(s)$ data.  Namely, the parameters
$a_k$ which parameterized his function were chosen to minimize
the $\chi^2$ defined as
\beq
\chi^2 = \sum_{i,j} \left[R_i - R_{fit}(s_i;a_k)
                    \right] V_{ij}^{-1}
                    \left[R_j - R_{fit}(s_j;a_k)
                    \right],
\eeq
where $R_i$ is the experimental value of the $i$-th measurement
at $s_i$, and $V^{-1} = (V_{ij}^{-1})$ is the inverse of the
variance--covariance matrix $V = (V_{ij})$:
\beq
V_{ij} = \langle \Delta R_i\,\Delta R_j
         \rangle.
\eeq
$\Delta R_i = R(s_i) - R_i$ is the difference 
between the true value of $R(s_i)$ and the measured
value $R_i$, and $\langle{\cal O}\rangle$
denotes the expectation value of ${\cal O}$.

\begin{figure}[t]
\begin{center}

\epsfig{file=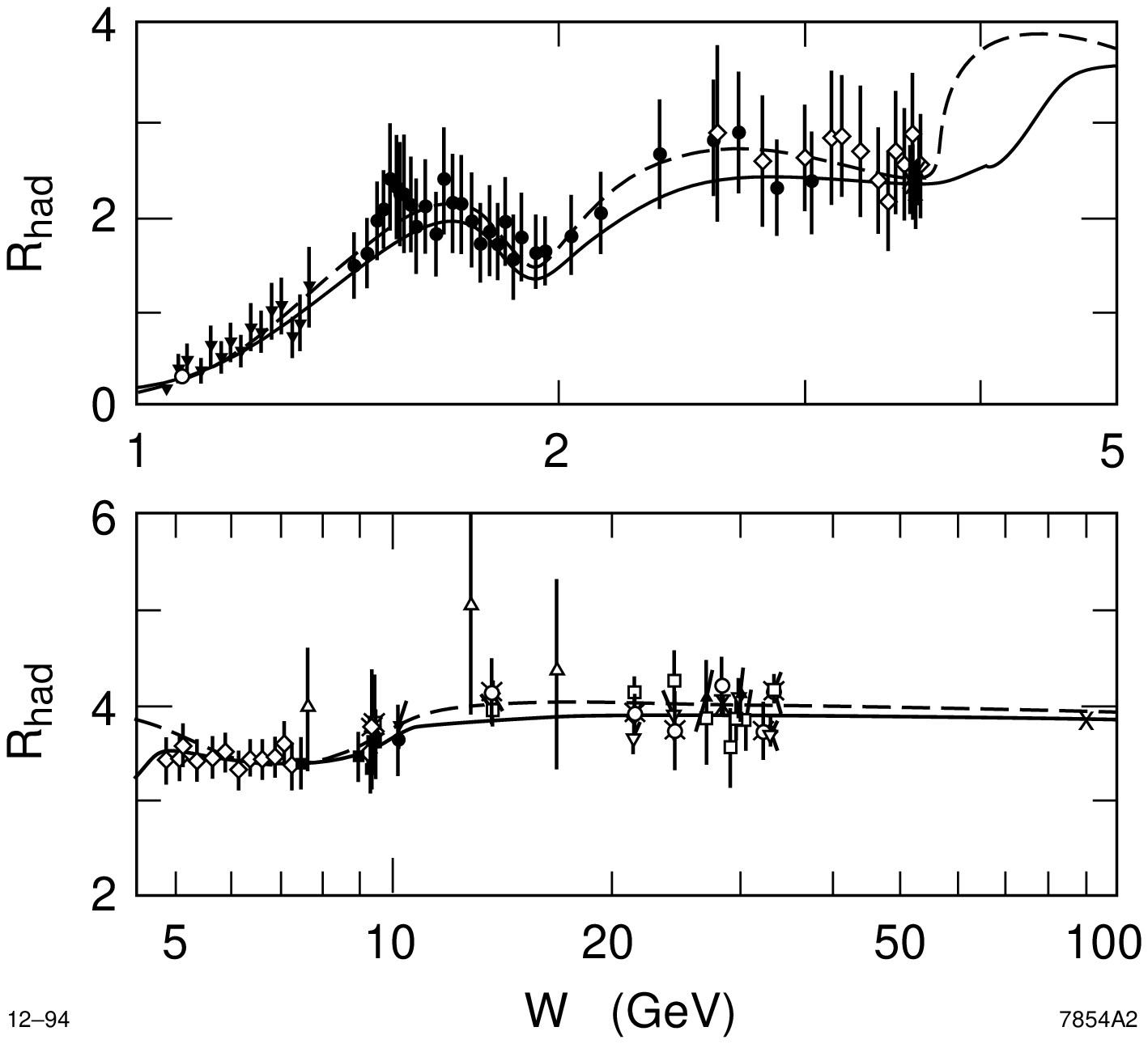,angle=0,height=2.5in}
\caption{ 
Fit to the continuum part of $R(s)$ with (solid line) and
without (broken line) correlations from normalization 
errors.  
The difference between the two in the 3.6--5 GeV
region is due to a different treatment of the higher
$\Psi$ resonances.
The error bars on the data points show the 
total (statistical+systematic) error.}
\label{SFIG1}

\epsfig{file=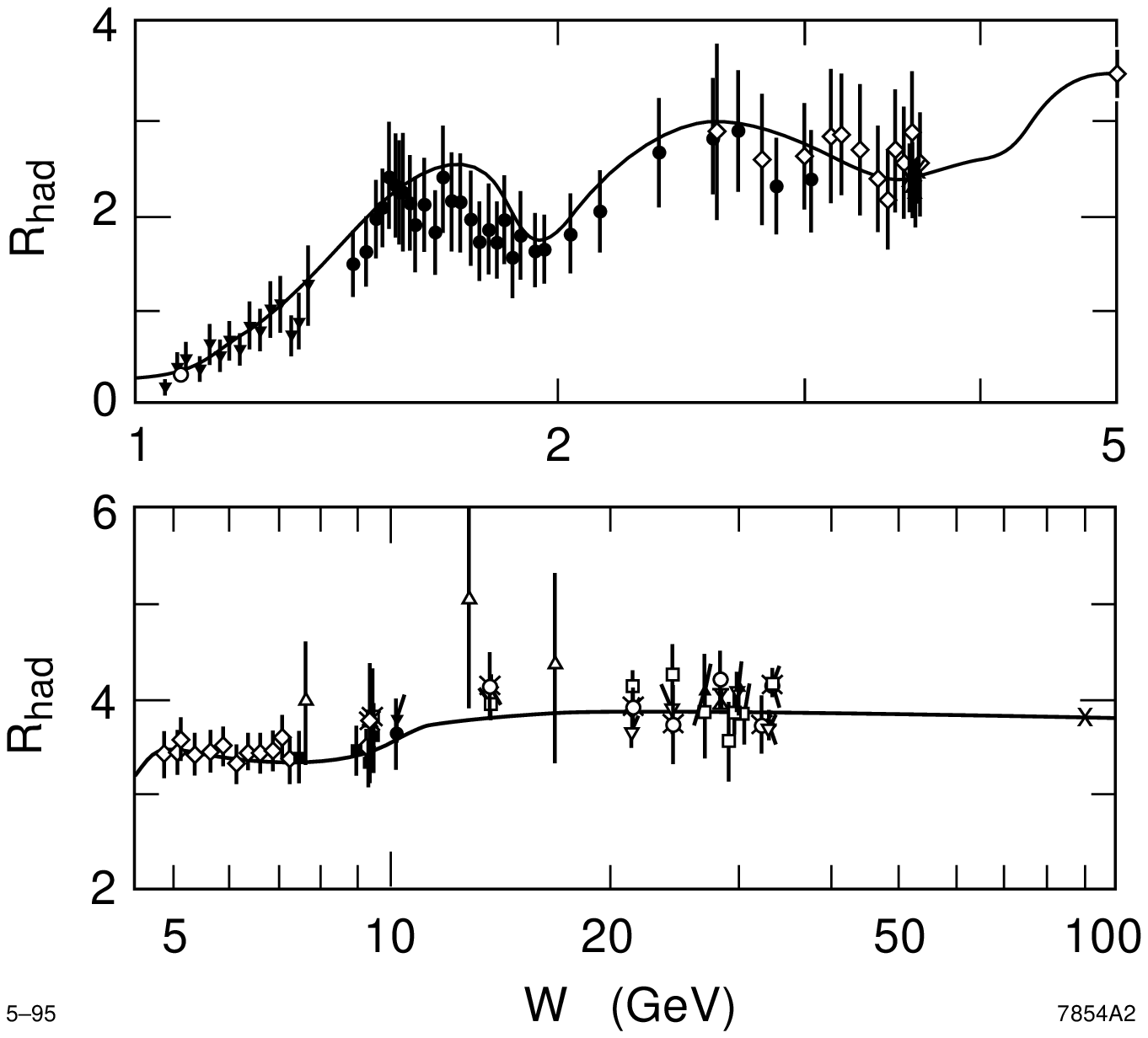,angle=0,height=2.5in}
\caption{The new fit by Swartz.
The Crystal Ball data point at $W = 3.670$ GeV is
hidden in the cluster of points in the vicinity.
The error bars show the total error.}
\label{SFIG2}

\end{center}
\end{figure}

Following Swartz,
I denote the uncorrelated point to point error on $R_i$
as $\sigma_i(ptp)$ and the correlated normalization error
as $\sigma_i(norm)$:
\beq
R(s_i) = R_i \pm \sigma_i(ptp) \pm \sigma_i(norm).
\eeq
Assuming that both Type~I and Type~II
correlations are 100\% when they are present, $V_{ij}$ is
given by
\beq
V_{ij} = \left\{
         \begin{array}{ll}
         \sigma_i^2(ptp) + \sigma_i^2(norm) & i=j \\
         \sigma_i(norm)\sigma_j(norm)       & 
             \mbox{$i\neq j$, with correlation } \\
         0                                  & 
             \mbox{$i\neq j$, no correlation}
         \end{array}
         \right.
\eeq
Using this expression for $V_{ij}$, Swartz obtained the fit shown 
with a solid line in Fig.~\ref{SFIG1}.  
For comparison, Fig.~\ref{SFIG1} also shows the fit
when all correlations are neglected with a broken line.

The difference between the two cases is significant.
When the uncorrelated fit was used, Swartz found that he reproduced
the result of Ref.~\citen{JEGER}, in good agreement with
Refs.~\citen{EJ,BP}, and the observation of Ref.~\citen{BJPV}
that the value of $\dahz$ is relatively insensitive 
to the interpolation method.
However, when he used the correlated fit, he found
\beq
\dahz = 0.02672 \pm 0.00075,
\eeq
and
\beq
\aiz = 129.07 \pm 0.10,
\label{SWARTZVALUE}
\eeq
which differed from Eq.~\ref{JEGERVALUE} by $0.20\approx 2\sigma$.

This analysis has been criticized\cite{EJ} on the grounds
that including normalization errors in the correlation matrix will
produce a fit which is biased towards smaller values of $R(s)$.
This effect is beautifully explained in a very nice 
paper by D'Agostini\cite{DAGOSTINI}.
Since I feel that this is a very important point which everyone
should know about, I have reproduced the basic argument in 
Appendix~A.

Swartz has subsequently updated his analysis to correct for
this problem and also to include some data\cite{CBALL2} which was
missing from his first analysis\cite{MORRIS2}.
The new fit to the continuum part of $R(s)$ is shown in Fig.~\ref{SFIG2},
and was obtained by minimizing
\beq
\chi^2 = \sum_{i}
         \frac{ [R_i - (1+\lambda_j\alpha_i)R_{fit}(s_i;a_k)]^2 }
              { \sigma_i^2(ptp) }
       + \sum_{j} \lambda_j^2,
\eeq
where $\alpha_i = \sigma_i(norm)/R_i$, and $\lambda_j$ are
normalization parameters which account for collective rescaling
of the data points in the $j$-th data set. (cf. Appendix~A.) 
Using this new fit, Swartz obtained
\beq
\dahz = 0.02752 \pm 0.00046,
\eeq
and
\beq
\aiz = 128.96 \pm 0.06.
\label{SWARTZVALUE2}
\eeq
The interesting thing about this result is that
only 1/3 of the difference between this value and the previous one
given in Eq.~\ref{SWARTZVALUE} is accounted for by the change from 
a biased to an unbiased fit.  
The remaining 2/3 of the shift is due to the inclusion of
a single data point at $W=3.670$ GeV from the Crystal Ball collaboration
\cite{CBALL2} which had a much smaller normalization error than all
other data points in the vicinity.
The effect of this point can be clearly seen by
comparing Fig.~\ref{SFIG2} and the uncorrelated fit in Fig.~\ref{SFIG1}.
Due to normalization correlations, 
the entire fit function below $W=3.670$ GeV has been pulled upwards 
and statistical fluctuations around this function suppressed.
This accounts for the larger value of $\dahz$ and smaller error.

\begin{figure}[t]
\begin{center}

\epsfig{file=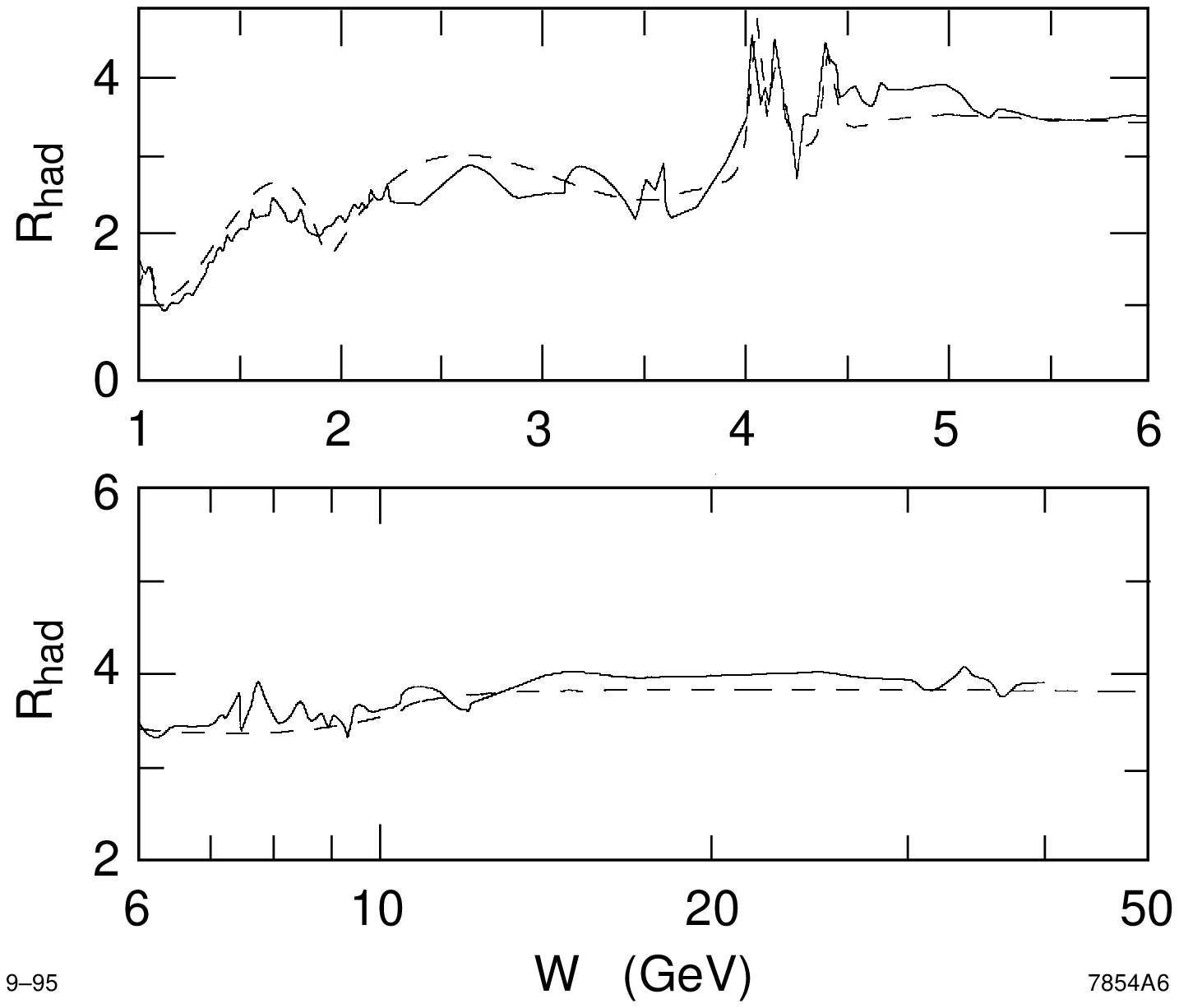,angle=0,height=2.5in}
\caption{Comparison of the new fit by Swartz (broken line)
and the trapezoidal integration of Eidelman and Jegerlehner
(solid line).}
\label{SFIG3}

\end{center}
\end{figure}

\subsection{Discussion}

In Refs.~\citen{BJPV,JEGER,EJ,BP} the difficulty of dealing with
Type~II correlations is discussed, but no strategy is presented as to how
they are actually dealt with.   Type~I correlations are not even
mentioned.   Given that the uncorrelated fit of Swartz\cite{MORRIS1}
reproduced the result of Refs.~\citen{JEGER,EJ,BP}, I believe it safe
to assume that none of these analyses took Type~I or Type~II 
correlations into account.
The independence of the result on the interpolation method found
in these works can also be understood as due to the neglect of
normalization correlations.

In the Swartz '95 analysis\cite{MORRIS2},
correlated data are collectively rescaled to obtain the best fit.
Both Type~I and Type~II correlations are considered.
The resulting fit function is quite different from the one integrated 
in the trapezoidal method of Ref.~\citen{EJ} as shown in Fig.~\ref{SFIG3}.
In my opinion, this analysis makes the best use of all the
information available in the experimental data.
However, the Swartz estimate is also susceptible to change
drastically by the inclusion of new data.
This is exemplified by the effect of the one data point
from Crystal Ball at $W=3.670$ GeV which `fixed' the global
normalization of $R(s)$ below this point.
If this point had not been included, then the estimate would have
been much closer to the Swartz '94 value\cite{MORRIS1}.
Every time a new and precise data point is included in the
analysis, the correlations will cause a ripple effect
which may shift the entire function upwards or downwards.
Therefore, to be on the conservative side, the error should
be doubled when using the Swartz estimate.
  
The Martin and Zeppenfeld value\cite{MARZEP} is obtained by
using perturbative QCD to rescale the experimental data, so if
one believes in perturbative QCD down to 3 GeV, then
this would be a good value to use.  One must remember, however,
that theoretical prejudice has entered into the $\az$ value.
The error is also artificially small due to use of perturbative
QCD and cannot be justified by the data alone.
To be on the conservative side, the error should be doubled 
in this case also.

My conclusion is that the Swartz '95 analysis with doubled
errors is probably the best estimate of $\az$ currently available.
Namely,
\beq
\aiz \approx 129.0 \pm 0.1.
\eeq
(Extra digits would be meaningless.)
A better determination of $\az$ is impossible with the
current quality of $R(s)$ data.   
Any further improvement requires
better measurements of $R(s)$ in the low energy region
below 10 GeV\cite{BP}.

%\newpage

\section{\boldmath The Value of $\asz$}

\subsection{Current status}

In the case of $\asz$, there exist more than one way to 
determine it's value.    
Refs.~\citen{HINCH,SHIFMAN} give comprehensive overviews 
of the many ways to measure $\asz$.
In table~\ref{aszlist}, I list some of the most recent 
determinations of $\asz$ using various techniques.

\begin{wraptable}{l}{7cm}
\caption{Current status of the determination of $\asz$.
}
\label{aszlist}
\begin{tabular}{|l|c|}
\hline
Method of Measurement
& $\asz$  \\
\hline\hline
Deep Inelastic Scattering\cite{SHIFMAN} 
& $0.112 \pm 0.005$      \\
\hline
GLS sum rule\cite{SHIFMAN}
& $0.108 \pm 0.008$      \\
\hline
$\Upsilon$ decay into 3 gluons\cite{YDECAY}
& $0.108 \pm 0.005$      \\ 
\hline
Voloshin\cite{VOLOSHIN}
& $0.109 \pm 0.001$      \\
\hline
Lattice\cite{MICHAEL}
& $0.112 \pm 0.007$    \\
\hline
\underline{LEP} &   \\
$Z$ lineshape data\cite{LEPDATA}
& $0.125 \pm 0.004$     \\
$R_\ell$ only\cite{LEPDATA}
& $0.126 \pm 0.005$     \\
event shapes and jet rates\cite{BETHKE}
& $0.123 \pm 0.006$     \\
\hline
\underline{SLD} &  \\
event shapes and jet rates\cite{SLD}
& $0.120 \pm 0.008$  \\
\hline
\end{tabular}
\end{wraptable}

As pointed out by Shifman\cite{SHIFMAN}, all the low energy
determinations of $\asz$ are clustered around $\asz = 0.11$
while all the high energy determinations at LEP/SLC
are clustered around $\asz = 0.125$.
This discrepancy, with a related discrepancy between the
SM prediction and experiment in $R_b$, can be interpreted as a sign of 
new physics as discussed in the talk by Hagiwara\cite{HAGIWARA}.

Another interesting possibility that was pointed out
by Consoli and Ferroni in Ref.~\citen{CONFER}
was that the error on the
LEP value of $R_\ell = \Gamma(Z\rightarrow hadrons)/
\Gamma(Z\rightarrow \ell^+\ell^-)$
may be grossly underestimated and that the actual error
may be roughly five times larger.
If that were the case, the error on the value of $\asz$
extracted from $R_\ell$ will be enhanced also by the
same factor and the discrepancy between the low energy 
and high energy determinations may vanish.

Unfortunately (or fortunately, depending on your point 
of view) their argument is flawed.
In the following, I will give an outline of 
the Consoli and Ferroni discussion and point out
where their logic is problematic.

\begin{table}[t]
\begin{center}

\caption{$R_\ell$ values from LEP 1994 data.}
\label{RellLEP}
\begin{tabular}{|l|llll|}
\hline
      & ALEPH & DELPHI & L3 & OPAL \\
\hline\hline
$R_e$    & $20.67\pm 0.13$ & $20.96\pm 0.16$
         & $20.94\pm 0.13$ & $20.90\pm 0.13$ \\
$R_\mu$  & $20.91\pm 0.14$ & $20.60\pm 0.12$ 
         & $20.93\pm 0.14$ & $20.855\pm 0.097$ \\
$R_\tau$ & $20.69\pm 0.12$ & $20.64\pm 0.16$
         & $20.70\pm 0.17$ & $20.91\pm 0.13$ \\
\hline
\end{tabular}

\end{center}
\end{table}

\begin{figure}[t]
\begin{center}
\epsfig{file=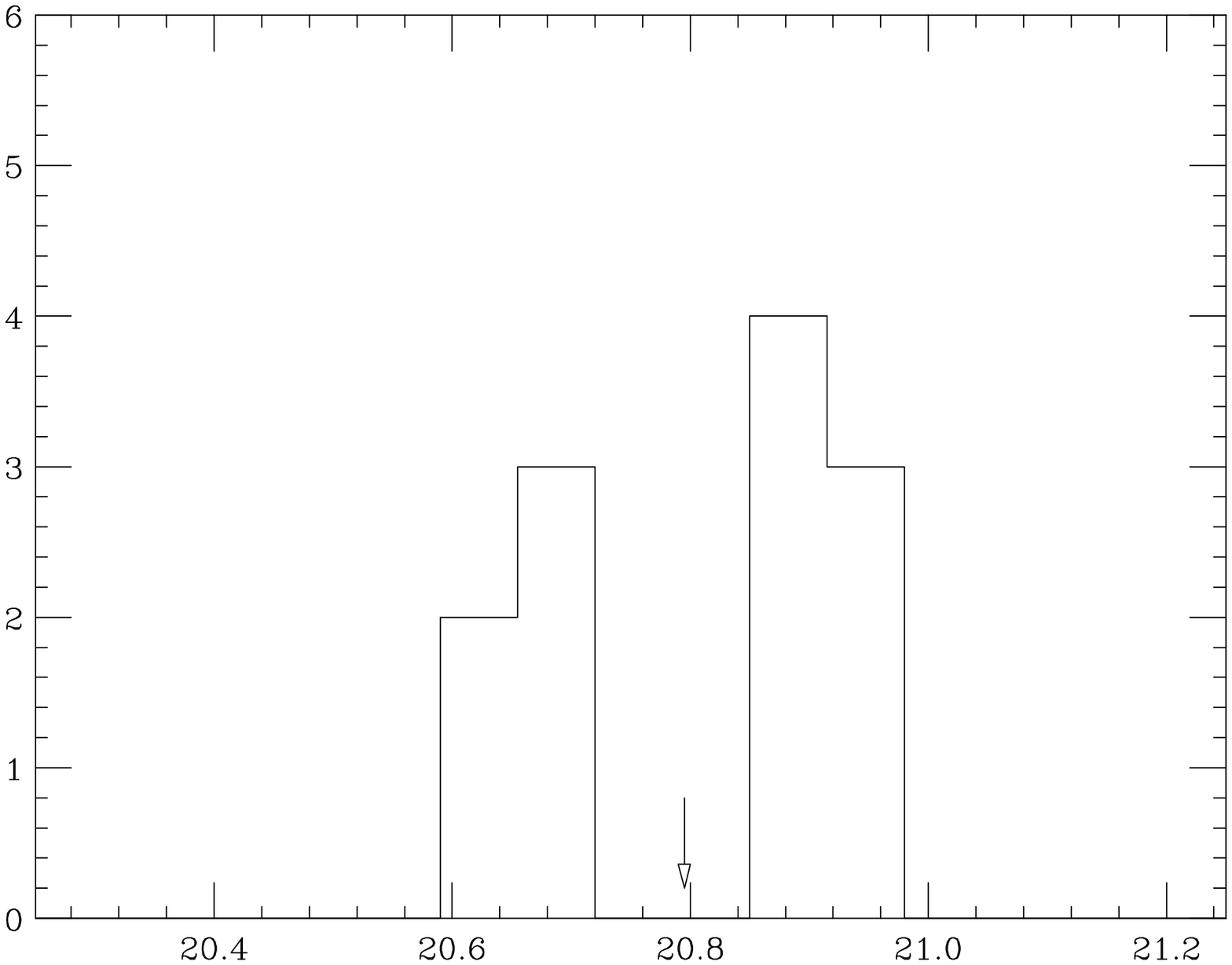,angle=90,height=2.0in}
\caption{The Distribution of LEP 1994 $R_\ell$ data.
         The arrow indicates the weighted average.}
\label{CFFIG}
%\end{center}
%\end{figure}
%
%\begin{figure}[t]
%\begin{center}
\epsfig{file=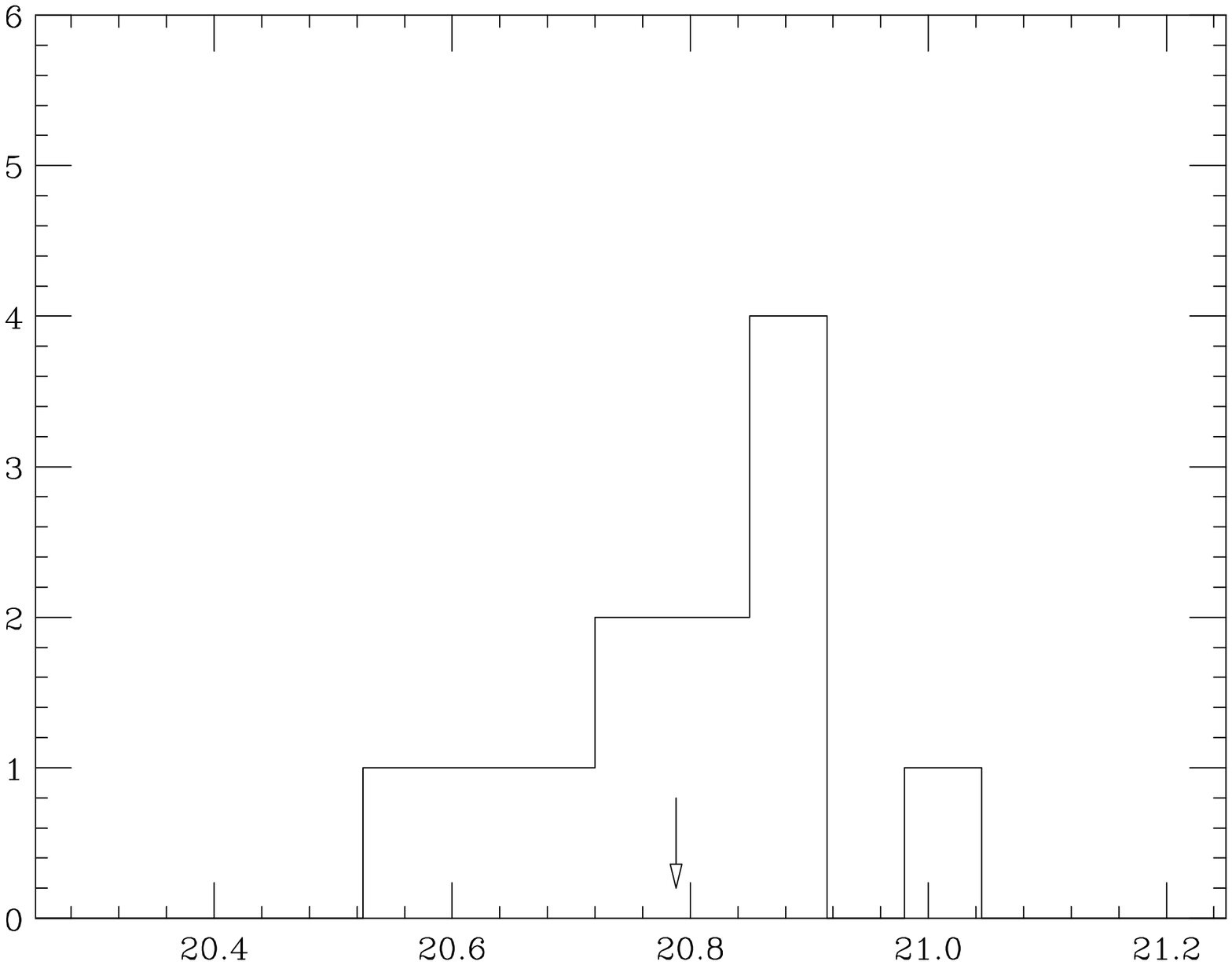,angle=90,height=2.0in}
\caption{The Distribution of LEP 1995 $R_\ell$ data.
         The arrow indicates the weighted average.}
\label{CFFIG2}
\end{center}
\end{figure}

\subsection{The Argument of Consoli and Ferroni}

The argument of Consoli and Ferroni\cite{CONFER}
was based on the $R_\ell$ values from the 1994 LEP data.
In Table.~\ref{RellLEP}, I list the data for
$R_e$, $R_\mu$, and $R_\tau$ from the four LEP experiments.
Neglecting small correlations and using the standard
error weighted average formula
\beq
\bar{x} = \frac{\sum_i\frac{x_i}{\sigma_i^2}}
               {\sum_i\frac{1  }{\sigma_i^2}},\qquad
\frac{1}{\bar{\sigma}^2}
        = \sum_i \frac{1}{\sigma_i^2},
\label{AVERAGE}
\eeq
the average of the 12 values listed in Table~\ref{RellLEP}
gave
\beq
R_\ell = 20.795 \pm 0.040
\label{RLEP}
\eeq
which corresponded to
\beq
\asz = 0.126 \pm 0.006.
\label{ASZLEP}
\eeq

The observation of Consoli and Ferroni was that when the
12 data points were plotted in a histogram, it looked like
Fig.~\ref{CFFIG} with two distinct peaks sandwiching the
average value.
This made them question whether the data points were normally
distributed around a common peak 
and whether it was correct to use Eq.~\ref{AVERAGE}.

In order to answer this question, they calculated 
a quantity called the {\it kurtosis} defined
by
\beq
\gamma = \mu_4/\mu_2^2 -3,
\eeq
where $\mu_n = \int (x-\bar{x})^n f(x) dx$ is the $n$-th moment
of the distribution, and found it to be $-1.8$.
If the LEP data points were normally distributed, 
the probability of the kurtosis $\gamma$ being this small is only
$4\times 10^{-4}$.  Therefore, they concluded that 
the LEP $R_\ell$ data was not normally distributed and that
Eq.~\ref{AVERAGE} could not be used.  

They argue that in the absence of a sensitive way to
combine the 12 measurements, the only thing that can be
said about the true value of $R_\ell$ is that it must lie
somewhere in the region 
\beq
20.60 \le R_\ell \le 20.98
\eeq
which is obtained by visual inspection of Fig.~\ref{CFFIG}.
This corresponds to
\beq
0.10 \le \asz \le 0.15
\eeq
which is perfectly consistent with the low energy value of
$\asz = 0.11$.

The problem with this argument is simple:
Eq.~\ref{AVERAGE} is actually valid 
{\it no matter what the distribution is} 
as long as the data points are unbiased
\footnote{By `unbiased' one means that $\vev{x_i} = x$.}
estimates of the true value.
\footnote{Determine the weights of a weighted average
$\bar{x} = \sum w_i x_i /\sum_i w_i$ so that 
$\bar\sigma^2=\vev{(x-\bar{x})^2}$
is minimized.  One always finds $w_i = 1/\sigma_i^2$ 
and $1/\bar\sigma^2 = \sum_i w_i$
without making any assumption about the distribution.}
The kurtosis is a quantity which measures how spread out the tails 
of the distribution is compared to the normal distribution, and the
fact that it is negative only means that the tails are less spread out
than normal.  (This is a typical symptom of LEP experiments
since data points that are too far off from the SM prediction
tend to get more attention and various corrections applied until
the agreement is improved.)   
The kurtosis cannot distinguish whether there is one peak or two.       

If there are really two peaks, the separation must be due
to systematic effects that are experiment and/or lepton flavor
dependent.  However, inspection of the numbers in Table~\ref{RellLEP}
shows that the data points for each experiment and each lepton 
flavor are evenly distributed between the two peaks.
There is no sign of any systematic effect so
the separation into two peaks for this particular data set must
have been purely statistical.   
This point is further supported by the latest LEP $R_\ell$ data\cite{LEPDATA}
in which the two peak structure has all but disappeared.
See Fig.~\ref{CFFIG2}.

Therefore, it is perfectly valid to use Eq.~\ref{AVERAGE} to
obtain Eq.~\ref{RLEP}, and hence Eq.~\ref{ASZLEP}, so the
discrepancy in the low and high energy determinations of $\asz$
still persists.

\section{Summary}

I have reviewed the current status of the determination of
$\az$ and $\asz$.

For $\az$, I reviewed all the recent evaluations of its
value and concluded that
\beq
\aiz = 129.0 \pm 0.1
\eeq
was probably the best estimate currently available.
Further improvements require better measurement of $R(s)$
at low energies below 10 GeV.

For the $\asz$, I reviewed the claim of Ref.~\citen{CONFER}
that the value $\asz$ determined from $R_\ell$ should
be assigned a larger error bar and concluded that the
argument was erroneous.

\section{Acknowledgements}

I would like to thank M.~L.~Swartz and D.~Zeppenfeld
for providing me with the postscript files for
Figs.~$2\sim 6$, and K.~Takeuchi for helpful discussions.
This work was supported in part by the United States Department of Energy
under Contract Number DE--AC02--76CH030000.

\appendix

\section{Correlations due to Normalization Errors}

In this appendix, I discuss the problem of performing a $\chi^2$
fit when normalization errors are present.  To avoid unnecessary
complications, I will only discuss the case of fitting a constant
$k$ to $n$ data points $x_i$ $(i=1,2,\cdots,n)$ 
which are all measurements of the same quantity $x$.
Extension to the case of fitting with a function is straightforward.

The standard method of fitting a constant $k$ to $n$ data points is
to minimize the $\chi^2$ defined as 
\beq
\chi^2 = \sum_{i,j}(x_i - k)V^{-1}_{ij}(x_j -k). 
\eeq
Here, $V^{-1}=(V^{-1}_{ij})$ is the inverse of the covariance matrix 
$V=(V_{ij})$ whose elements are given by:
\beq
V_{ij} = \vev{\epsilon_i\epsilon_j},
\eeq
where $\vev{\cal O}$ denotes the expectation value of ${\cal O}$
and $\epsilon_i$ is the difference between the
the true value of $x$ and the measured value $x_i$:
\beq
x = x_i + \epsilon_i.
\eeq
It is assumed that the $x_i$'s are unbiased measurements of 
$x$, and that the $\sigma_i$'s 
correctly estimate the expectation value 
of $\epsilon_i^2$, {\sl i.e.}
\beq
\vev{\epsilon_i}   = 0,\qquad 
\vev{\epsilon_i^2} = \sigma_i^2.
\eeq
When none of the $\epsilon_i$'s are correlated, the covariance
matrix $V$ and its inverse are diagonal:
\beq
V_{ij}      = \sigma_i^2 \delta_{ij},\qquad
V^{-1}_{ij} = \frac{\delta_{ij}}{\sigma_i^2}
\eeq
and we find that the above definition of $\chi^2$ reduces to
the familiar 
\beq
\chi^2 = \sum_i \left( \frac{x_i - k}{\sigma_i}
                \right)^2,
\eeq
which is minimized at
\beq
k = \frac{ \displaystyle \sum_i\frac{x_i}{\sigma_i^2} }
         { \displaystyle \sum_i\frac{ 1 }{\sigma_i^2} }
  \equiv \bar{x}.
\eeq
Since $\vev{\bar{x}}=x$, we have an unbiased estimate of the
true value $x$.

In order to apply this formalism to the case where 
there is an overall normalization uncertainty 
in addition to uncorrelated point to point errors,
I introduce a scale factor $\f = 1 + \epsilon_\f$,
where $\vev{\epsilon_\f}=0$, and
$\vev{\epsilon_\f^2} = \sigma_\f^2$, 
and modify the relation between $x$ and $x_i$ to
\beq
x = \f (x_i + \epsilon_i)
  = x_i + \epsilon_i + \epsilon_\f x_i + O(\epsilon^2)
\eeq
The correlation matrix $V$ is then modified to
\beq
V_{ij}   =   \vev{(\epsilon_i+\epsilon_\f x_i)
                  (\epsilon_j+\epsilon_\f x_j)
                 }
       \;=\; \sigma_i^2\delta_{ij} + \sigma_\f^2 x_i x_j
\eeq
and its inverse to
\beq
V^{-1}_{ij} 
= \frac{\delta_{ij}}{\sigma_i^2}
- \frac{1}{\left( \displaystyle 
                  \frac{1}{\sigma_\f^2}
                 +\sum_{k}\frac{x_k^2}{\sigma_k^2}
           \right)
          }
              \frac{x_i}{\sigma_i^2}
              \frac{x_j}{\sigma_j^2}.
\eeq
The second term comes from the correlation due to the 
overall normalization error and goes to zero in the limit
$\sigma_\f \rightarrow 0$.  The $\chi^2$ is then
\beq
\chi^2 
= \sum_i \left( \frac{x_i - k}{\sigma_i}
         \right)^2
- \frac{1}{\left( \displaystyle 
                  \frac{1}{\sigma_\f^2}
                 +\sum_{k}\frac{x_k^2}{\sigma_k^2}
           \right)
          }
  \left[ \sum_i \frac{(x_i - k)x_i}{\sigma_i^2}
  \right]^2.
\label{CHISQ2}
\eeq
which is minimized at
\beq
k = \frac{\bar{x}}
         {\displaystyle
          1+\sigma_\f^2 \sum_i 
            \left( \frac{x_i - \bar{x}}{\sigma_i} 
            \right)^2 
         }
\equiv \bar{k}.
\eeq
The expectation value of this quantity is 
\beq
\vev{\bar{k}} = \frac{x}{1 + (n-1)\sigma_f^2}
\eeq
so it is clearly biased towards values smaller than
the true value $x$.  In fact, $\vev{\bar{k}}\rightarrow 0$
in the limit $n\rightarrow \infty$.

The source of this bias can be understood as follows.
It is easy to show that minimizing the $\chi^2$ defined in
Eq.~\ref{CHISQ2} is equivalent to minimizing
\beq
\chi^2 = \sum_i \left( \frac{\f x_i - k}
                            {\sigma_i}
                \right)^2
       + \left( \frac{\f-1}{\sigma_\f}
         \right)^2.
\label{CHISQ3}
\eeq
Obviously, the first summation term is minimized at 
$\f = k = 0$, while the second term is minimized at $f=1$.
Therefore, as the number of data points
increases, minimizing this $\chi^2$ has the effect of 
decreasing the first term by decreasing $\f$
at the expense of making the second term large.

The problem with
Eq.~\ref{CHISQ3}
is that only the data points $x_i$ rescale with
$\f$ and the errors $\sigma_i$ are left untouched.
But since $\f$ is a normalization factor multiplying the
errors also, they too should scale with $\f$.
Otherwise, the $\chi^2$ improves the agreement
between data points by simply rescaling everything to zero!
 
This problem can be corrected by defining the $\chi^2$ as
\beq
\chi^2 = \sum_i \left( \frac{ x_i - k/\f }
                            { \sigma_i }
                \right)^2
       + \left( \frac{\f-1}{\sigma_\f}
         \right)^2. 
\eeq
In this case, the result is rather trivial since this $\chi^2$
is minimized at $f=1$, $k=\bar{x}$.
However, this definition can be easily extended to cases where
there is more than one data set, each with its respective
overall normalization error.  In the extended cases, finding
the minimum becomes a non--linear problem which is 
best solved by computer.

\end{document}